\documentclass[twocolumn]{aastex701}

\usepackage{booktabs}
\usepackage{float}
\usepackage{mathtools}
\usepackage{amsmath}
\usepackage{cleveref}
\usepackage{comment}
\usepackage{upgreek}
\usepackage{xcolor}
\usepackage[nolist]{acronym}
\usepackage{subfigure}
\usepackage{tabularx}
\usepackage{siunitx}
\usepackage{booktabs}
\usepackage{makecell}
\usepackage{longtable}
\usepackage{caption}
\usepackage{adjustbox}

\definecolor{linkblue}{HTML}{0645AD}
\hypersetup{
    colorlinks=true,
    urlcolor=linkblue
}

\definecolor{vg}{rgb}{0.0, 0.0, 0.9}

\definecolor{pj}{rgb}{0.9, 0.0, 0.0}

\begin{document}

\title{High Frequency Wideband Study of FRB 20240114A with the Allen Telescope Array}
\shorttitle{FRB 20240114A wideband study - Allen Telescope Array}

\correspondingauthor{Param Joshi}
\email{paramjoshi1812@gmail.com}

\shortauthors{Joshi et al.}

\author[0009-0005-9293-0655]{Param Joshi}
\affiliation{SETI Institute, 339 Bernardo Ave, Suite 200 Mountain View, CA 94043, USA}
\affiliation{Department of Physics and Astronomy, National Institute of Technology, Rourkela, 769008, India}
\email{paramjoshi1812@gmail.com}

\author[0000-0002-8604-106X]{Vishal Gajjar}
\affiliation{SETI Institute, 339 Bernardo Ave, Suite 200 Mountain View, CA 94043, USA}
\affiliation{Department of Astronomy, University of California Berkeley, Berkeley, CA 94720, USA}
\email{vgajjar@seti.org}

\author[0000-0001-5915-1127]{Joel Earwicker}
\affiliation{SETI Institute, 339 Bernardo Ave, Suite 200 Mountain View, CA 94043, USA}
\email{jearwicker@seti.org}

\author[0000-0001-7057-4999]{Sofia Z. Sheikh}
\affiliation{SETI Institute, 339 Bernardo Ave, Suite 200 Mountain View, CA 94043, USA}
\affiliation{Berkeley SETI Research Center, 339 Campbell, Berkeley, CA 947206}
\email{ssheikh@berkeley.edu}

\author[0000-0003-1051-282X]{Mohammed A. Chamma}
\affiliation{Department of Physics, McGill University, 3600 rue University, Montréal, QC H3A 2T8, Canada}
\affiliation{Trottier Space Institute, McGill University, 3550 rue University, Montréal, QC H3A 2A7, Canada}
\affiliation{Department of Physics and Astronomy, The University of Western Ontario, 1151 Richmond Street, London, Ontario N6A 3K7, Canada}
\email{mchamma@uwo.ca}

\author[0000-0002-7735-5796]{Joe Bright}
\affiliation{SETI Institute, 339 Bernardo Ave, Suite 200 Mountain View, CA 94043, USA}
\affiliation{Breakthrough Listen, University of Oxford, Denys Wilkinson Building, Keble Road, Oxford OX1 3RH, UK}
\email{jbright@seti.org}

\author[0000-0001-5576-2254]{Luigi F. Cruz}
\affiliation{SETI Institute, 339 Bernardo Ave, Suite 200 Mountain View, CA 94043, USA}
\email{luigifcruz@gmail.com}

\author[0009-0007-4933-465X]{Roy H. Davis}
\affiliation{SETI Institute, 339 Bernardo Ave, Suite 200 Mountain View, CA 94043, USA}
\email{WB9RKN@san.rr.com}

\author[0000-0003-3197-2294]{David R. DeBoer}
\affiliation{SETI Institute, 339 Bernardo Ave, Suite 200 Mountain View, CA 94043, USA}
\affiliation{Department of Astronomy, University of California Berkeley, Berkeley, CA 94720, USA}
\email{ddeboer@berkeley.edu}

\author[0009-0001-8677-372X]{R. A. Donnachie}
\affiliation{SETI Institute, 339 Bernardo Ave, Suite 200 Mountain View, CA 94043, USA}
\email{radonnachie@gmail.com}

\author[0000-0002-0161-7243]{Wael Farah}
\affiliation{SETI Institute, 339 Bernardo Ave, Suite 200 Mountain View, CA 94043, USA}
\email{wael.a.farah@gmail.com}

\author[]{Phil Karn}
\affiliation{SETI Institute, 339 Bernardo Ave, Suite 200 Mountain View, CA 94043, USA}
\email{karn@ka9q.net}

\author[0000-0003-4117-953X]{João Paolo C. M. Oliveira}
\affiliation{SETI Institute, 339 Bernardo Ave, Suite 200 Mountain View, CA 94043, USA}
\email{paolocmo@gmail.com}

\author[0000-0002-6341-4548]{Karen I. Perez}
\affiliation{SETI Institute, 339 Bernardo Ave, Suite 200 Mountain View, CA 94043, USA}
\email{kperez@seti.org}

\author[0000-0002-3430-7671]{Alexander W. Pollak}
\affiliation{SETI Institute, 339 Bernardo Ave, Suite 200 Mountain View, CA 94043, USA}
\affiliation{Department of Astronomy, University of California Berkeley, Berkeley, CA 94720, USA}
\email{apollak@seti.org}

\author[0000-0003-2828-7720]{Andrew Siemion}
\affiliation{SETI Institute, 339 Bernardo Ave, Suite 200 Mountain View, CA 94043, USA}
\affiliation{Department of Astronomy, University of California Berkeley, Berkeley, CA 94720, USA}
\affiliation{Sub-Department of Astrophysics, University of Oxford, Banbury Road, OX1 3RH, UK}
\affiliation{Jodrell Bank Centre for Astrophysics (JBCA), Department of Physics and Astronomy, Alan Turing Building, The University of Manchester, M13 9PL, UK}
\affiliation{Department of Physics, Faculty of Science, University of Malta, Msida MSD 2080, Malta}
\email{siemion@berkeley.edu}

\author[]{Michael Snodgrass}
\affiliation{SETI Institute, 339 Bernardo Ave, Suite 200 Mountain View, CA 94043, USA}
\email{mlsnodgrass@verizon.net}

\begin{abstract}
We present a high-frequency, wideband observing campaign of the hyperactive repeating fast radio burst FRB\,20240114A with the Allen Telescope Array. Between 27 January and 29 October 2024, we obtained 1167\,hr of on-source observations across 1344\,MHz of simultaneous bandwidth covering frequencies from approximately 900\,MHz to 7620\,MHz. We detected 97 bursts between $\sim900$\,MHz and $\sim5$\,GHz, including a strong S-band activity episode, while no bursts were detected in the highest-frequency tunings above $\sim5$\,GHz despite substantial exposure. This campaign provides one of the very few extended samples of repeating-FRB activity above 3\,GHz, a regime that remains sparsely sampled. We find that the burst rate varies strongly with both observing frequency and epoch, confirming that the emission from FRB\,20240114A is highly chromatic and band-limited. We measure the spectro-temporal properties of the bursts and their sub-components, confirming that fractional bandwidth remains approximately scale-invariant. Sub-burst durations decrease toward higher frequencies, and the magnitude of the downward drift rate increases with frequency. The cumulative spectral-energy-density distribution above our completeness threshold is well described by a shallow power law, indicating that high-energy bursts contribute substantially to the observed energy output. We also compare our detections with recently proposed long-timescale frequency-modulation models and find that the ATA high-frequency burst storm is not consistent with a strictly phase-coherent modulation inferred from other datasets. Our results demonstrate that incomplete time-frequency coverage can bias interpretations of burst activity and highlight the need for sustained, simultaneous wideband monitoring of hyperactive repeaters.
\end{abstract}

\keywords{\href{http://astrothesaurus.org/uat/2008}{Radio transient sources (2008)}; \href{http://astrothesaurus.org/uat/1339}{Radio bursts (1339)}}

\section{Introduction} 
\label{sec:intro}

Fast Radio Bursts (FRBs; \cite{lorimer_bright_2007}) are millisecond-duration, highly energetic radio transients of extragalactic origin. Repeating FRBs have now been detected across more than two orders of magnitude in frequency, from 110\,MHz \citep{pleunis_lofar_2021} to 8\,GHz \citep{gajjar_highest_2018}, and their burst rates, spectra, and spectro-temporal morphology often vary strongly with both frequency and epoch. Understanding this chromatic and temporal variability is central to distinguishing intrinsic emission physics from propagation and instrumental selection effects, and requires sustained, wideband monitoring that overlaps in time and frequency across multiple facilities.

FRB\,20240114A is among the brightest and most active repeaters discovered to date \citep{CHIME_FRB_2024,CHIME_FRB_2025}. Within weeks of the CHIME/FRB announcement, it was followed up globally, with detections reported from 327\,MHz \citep{Westerbork_Atel} through 6\,GHz \citep{limaye2026,Effelsberg_ATel}. Long campaigns with FAST \citep{FAST_main_paper,FAST_ATel}, Parkes/Murriyang \citep{Uttarkar2026,Parkes_Atel}, Effelsberg \citep{limaye2026}, and Tianma \citep{shangai_longterm_paper} have established that the source is broadband yet band-limited in any single epoch, with burst activity that evolves in both time and frequency. \citet{Uttarkar2026} find that the mean emission frequency rises secularly over 2024, while \citet{energydist_paper} report a flattened high-energy tail in the burst-energy distribution and \citet{FAST_drifting_paper3} identify predominantly downward-drifting ``sad-trombone'' burst clusters. Two independent periodicities have recently been claimed: a $\sim$143.4\,day cycle in burst \emph{activity} from FAST \citep{FAST_periodicity_paper2} and a $\sim$112.9\,day cycle in the \emph{central emission frequency} from a Parkes/Murriyang sample analysed by \citet{Li2026_freqperiod}. However, these findings rely on datasets that differ markedly in their frequency coverage, bandwidth, sensitivity, and observing cadence. In particular, the Parkes Ultra-Wideband receiver provided excellent sampling but was limited to frequencies below 3.9\,GHz \citep{Uttarkar2026}, whereas the Effelsberg campaign achieved an exceptionally broad continuous bandwidth from 1.3 to 6\,GHz, but its observations were collected more sparsely in time \citep{limaye2026}.

These gaps motivate high-frequency, wideband monitoring that can sample both the spectral and temporal evolution of FRB\,20240114A. Proposed central-frequency modulations must be tested against independent observations with different bandpasses and cadences, since incomplete frequency--time coverage can miss spectrally localized activity episodes and bias the inferred evolution. The Allen Telescope Array (ATA) is well suited to this task: its log-periodic feeds and independently tunable Local Oscillator (LO) chains provide $\sim$2.8\,GHz of continuous bandwidth per observation and allow coverage from $\sim$900\,MHz to $>$10\,GHz. Following our first report of wideband emission above 2\,GHz \citep{ATA_Atel}, we carried out an extended campaign between 27 January and 29 October 2024, stepping through ten LO tunings across the ATA band. In this paper, we describe the observing strategy, search pipeline, and data-reduction methods; measure the burst spectro-temporal and energy properties with \texttt{FRBGUI} \citep{Chamma_2023_FRBGUI}; and compare the ATA sample with contemporaneous Effelsberg, Parkes/Murriyang, FAST, and Tianma datasets to assess the source's broadband behavior and recent periodicity claims.

The paper is organized as follows. Section~\ref{sec:observation} describes the ATA instrument, observing strategy, and single-pulse search pipeline. Section~\ref{sec:burst_properties} summarizes data preparation, flux calibration, and \texttt{FRBGUI}-based property extraction. Section~\ref{sec:analysis} presents the frequency-dependent activity and periodicity assessment, spectro-temporal scaling with frequency, isotropic energy, cumulative Spectral Energy Density (SED) distributions, and the drift-rate--duration relation. We conclude in Section~\ref{sec:conclusion}.

\section{Observations}
\label{sec:observation}

\subsection{The Allen Telescope Array: Instrument Specifications}
\label{ssec:ATA_specifications}

The Allen Telescope Array (ATA) is a 42-element interferometer consisting of 6.1-m dishes located at Hat Creek Radio Observatory in Northern California, owned and operated by the SETI Institute, Mountain View, CA. Each fully steerable, offset Gregorian dish is equipped with a refurbished ``Antonio'' log-periodic feed, which is dual-polarization and provides wide instantaneous frequency coverage from 1--11\,GHz. The feeds are housed in cryostats and cooled to approximately 70\,K. Analog signals are amplified and transmitted over optical fiber, after which each antenna signal is split and mixed to produce up to four independently tunable signal chains, denoted as LO tunings `a', `b', `c', and `d'. Each tuning covers approximately 672\,MHz. The ATA backend can be configured for different science cases by selecting parameters such as the correlator integration time, the number of beams formed by the beamformer, and the desired polarization products. For the observations reported in this work, we used two consecutive tunings covering total 1344\,MHz of continuous bandwidth in beamformer mode. Further details on the digital signal processing chain and instrumental configuration are given in \cite{Sofia_FRB2022_paper}, which closely matches the configuration used for the observations presented here.

\subsection{Observations}
\label{ssec:Observations}
\begin{figure*}
    \centering
    \includegraphics[width=0.7\linewidth]{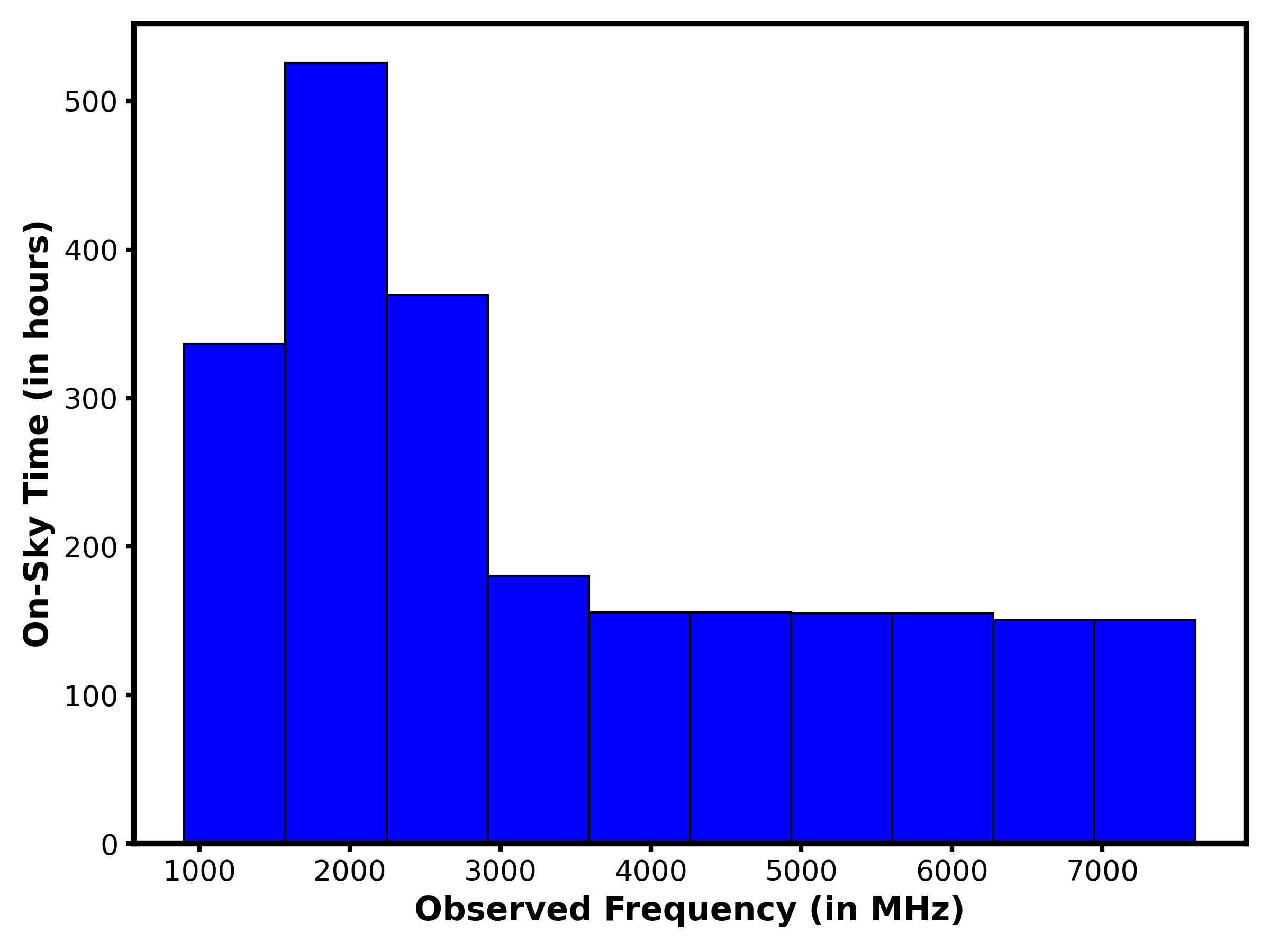}
    \caption{Total on-sky observing time for FRB\,20240114A as a function of observed frequency. The ATA campaign used multiple LO tunings spanning approximately 900–7620 MHz, with the largest exposure obtained at the lower-frequency tunings. The observing time shown here includes all data acquired between 27 January and 29 October 2024.}
    \label{fig:observation_plot}
\end{figure*}

\begin{figure*}
    \centering
    \includegraphics[width=0.8\linewidth]{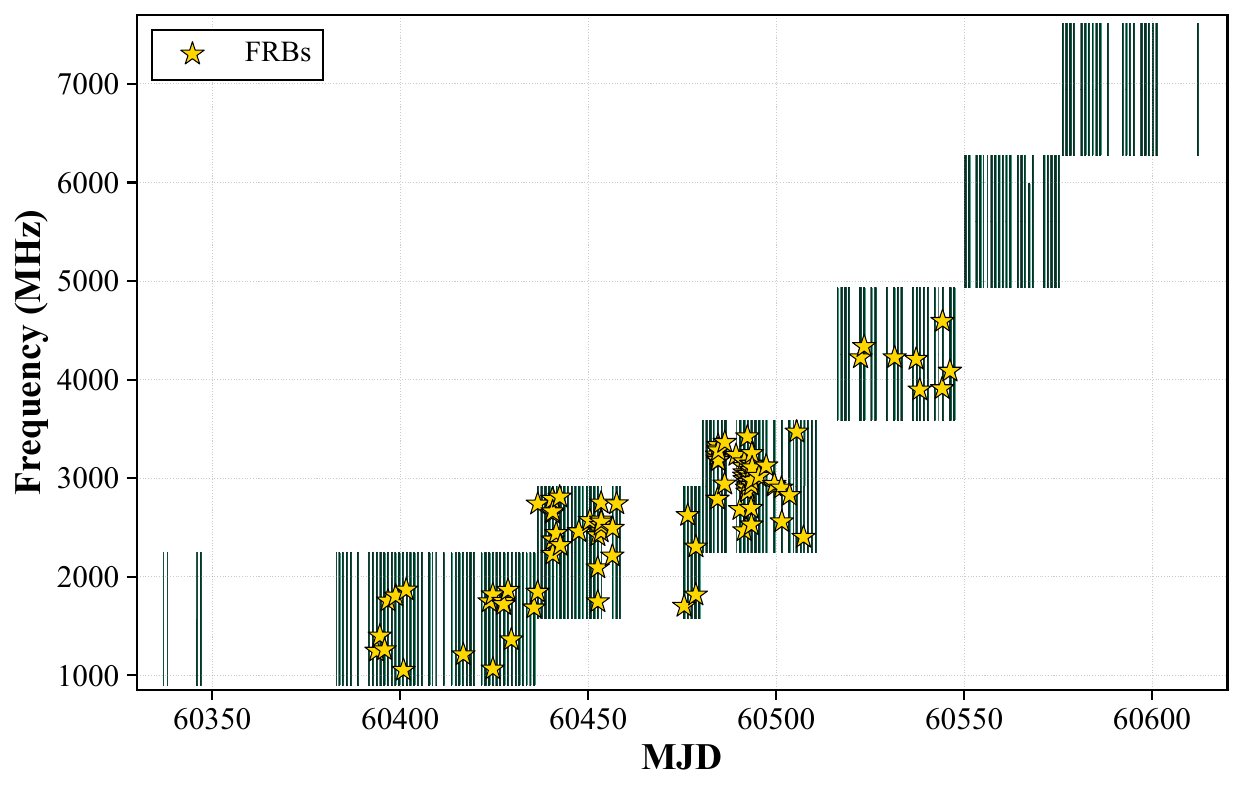}
    \caption{Temporal and frequency coverage of the ATA observations of FRB\,20240114A. Dark green vertical segments show the 1344\,MHz frequency range (two consecutive LOs) covered by each observing scan as a function of MJD, while yellow stars mark the detected FRB bursts. The observations span approximately 900--7620\,MHz and show the progression from lower-frequency tunings early in the campaign to higher-frequency tunings at later epochs.}
    \label{fig:frequency_vs_MJD}
\end{figure*}

We began observations of FRB\,20240114A with the ATA on 27 January 2024, one day after the CHIME/FRB Collaboration announced its detection on 26 January 2024 \citep{CHIME_FRB_2024}. The initial observing campaign consisted of 1-hour observing sessions, totaling 18 hours distributed non-uniformly over 12 days between 27 January and 7 February 2024. Following a gap caused by ongoing maintenance work, observations resumed on 13 March 2024. In this phase, we observed the source for 5 hours per day, typically divided into 30-minute scan blocks. In total, we accumulated 1167 hours of observations of FRB\,20240114A through 29 October 2024. The campaign began with a tuning centered at 1236\,MHz and eventually covered 10 unique LO tunings, extending up to a center frequency of 7284\,MHz. Overall, these tunings span a frequency range from approximately 900\,MHz to 7620\,MHz. The total observing time obtained at each tuning is shown in Figure~\ref{fig:observation_plot}. Progression of these LO tunings with MJD is shown in Figure \ref{fig:frequency_vs_MJD}. 

\begin{table*}[ht]
\centering
\begin{tabular}{|c|
                >{\centering\arraybackslash}p{0.13\linewidth}|
                >{\centering\arraybackslash}p{0.1\linewidth}|
                >{\centering\arraybackslash}p{0.1\linewidth}|
                >{\centering\arraybackslash}p{0.1\linewidth}|
                >{\centering\arraybackslash}p{0.1\linewidth}|
                >{\centering\arraybackslash}p{0.13\linewidth}|}
\hline
\textbf{MJD} & 
\textbf{Frequency Range (in MHz)} & 
\textbf{Number of antennas} & 
\textbf{No. of hours observed} & 
\textbf{Minimum Fluence (Jy-ms)} & 
\textbf{Number of bursts detected} & 
\textbf{Burst Rate (number of bursts hr$^{-1}\times10^{-2}$)} \\ \hline

60336 -- 60435 & 900--2244   & 20 & 336.4 & 1.85  & 16 & 4.75 \\ \hline
60436 -- 60458 & 1572--2916  & 20 & 148.32 & 2.11 & 24 & 16.18 \\ \hline
60475 -- 60479 & 1572--2916  & 28 & 41 & 1.59  & 4  & 9.75 \\ \hline
60480 -- 60510 & 2244--3588  & 28 & 180.16 & 1.77  & 45 & 24.97 \\ \hline
60516 -- 60547 & 3588--4932  & 28 & 155.62 & 2.33  & 8  & 5.14 \\ \hline
60550 -- 60575 & 4932--6276  & 28 & 155 & 2.12 & -- & -- \\ \hline
60576 -- 60612 & 6276--7620  & 28 & 150.5 & 2.29 & -- & -- \\ \hline

\end{tabular}
\caption{Summary of the ATA observing campaign for FRB\,20240114A, grouped by MJD range, frequency coverage, array configuration, and burst yield. The listed frequency ranges correspond to the combined bandwidth from the two simultaneous LO tunings used in each scan. During the early part of the campaign, several observing sessions used overlapping LO pairs, resulting in repeated coverage near 2\,GHz. At later epochs, the LO tunings were stepped to contiguous, non-overlapping frequency ranges to extend the search coverage up to 7620\,MHz.}
\label{tab:obs_table}
\end{table*}

From the beginning of the observing campaign through 28 May 2024, the ATA observations used 20 antennas, yielding 484\,hr of on-source data in this configuration. After a short interruption for software and hardware development, observations resumed on 14 June 2024 with an expanded 28-antenna array. This configuration was used for the remainder of the campaign, contributing 683\,hr of on-source data and providing improved sensitivity to single bursts. During the initial phase of the campaign, the telescope was pointed using the J2000 coordinates reported in the CHIME/FRB discovery announcement \citep{CHIME_FRB_2024}. Beginning on 29 March 2024, we adopted the improved EVN localization reported in \citep{EVN_localisation_ATel}. As a result, 97\,hr of on-source data were obtained using the original CHIME/FRB position prior to the availability of the refined localization.

Before each observing session, we observed a standard flux and phase calibrator, either 3C286 or 3C147, in configuring the array in an imaging mode. These observations were used to evaluate the system sensitivity and to derive updated delay and phase solutions for instrumental calibration. We also included regular observations of the bright pulsar J0332+5434 to validate the system performance and to test the SPANDAK single-pulse detection pipeline.

\subsection{FRB Searching with SPANDAK} 
\label{ssec:FRB_Searching}
For the observations described in the previous section, the ATA beamformer produced Stokes-I, 32-bit SIGPROC filterbank files\footnote{\href{https://sigproc.sourceforge.net/sigproc.pdf}{https://sigproc.sourceforge.net/sigproc.pdf}} with frequency and time resolutions of 0.5\,MHz and $64\,\upmu\textrm{s}$, respectively. These data were recorded for offline FRB searches. 
To expedite processing and ensure compatibility with the previous version of \texttt{HEIMDALL}  \citep{Barsdell_2012_heimdall} input format, the native 32-bit filterbank files were converted to 8-bit resolution before searching. For each observation, data from the seven compute nodes were spliced together separately for each of the two simultaneous LO tunings. This produced two 8-bit filterbank files per scan, each covering one full 672\,MHz tuning, corresponding to LO~a and LO~b.

These spliced filterbank products were searched using \texttt{SPANDAK} \citep{gajjar_highest_2018}, a single-pulse search and candidate-vetting pipeline built around \texttt{HEIMDALL}, a GPU-accelerated code for searching dispersed radio transients \citep{Barsdell_2012_heimdall}. The candidates produced by \texttt{HEIMDALL} were subsequently filtered and organized by \texttt{SPANDAK}, which generates candidate tables and diagnostic plots for visual inspection. In addition, we used a Convolutional Neural Network based machine-learning classifier, which was trained on ATA data and previously detected FRB candidates (see \citealt{Gajjar_2022_ML} for more details). This classifier was used to reduce the number of false-positive candidates requiring manual inspection. Candidates that passed these stages were assigned to category `A' and retained for final visual inspection.

Prior to searching for FRBs, known persistent RFI-affected channels were masked during the \texttt{SPANDAK} searches for most of the data obtained after the start of the S-band observations. At higher observing frequencies, where the data were affected by strong zero-DM RFI, we additionally applied zero-DM filtering before searching. These mitigation steps substantially reduced the number of false-positive candidates and the manual inspection load. The search was performed over a DM range of 500--550\,pc\,cm$^{-3}$. The signal-to-noise ratio (S/N) threshold was set to 10, and the maximum number of candidates per second was set to 15, based on the RFI environment and the sensitivity of the observations. The maximum boxcar width was set to 1024 samples, corresponding to a maximum searched pulse width of approximately 65\,ms. All other search parameters were left at their default values with the \texttt{HEIMDALL} tool. The search was performed independently on each spliced filterbank file, corresponding to a single LO tuning with a fixed bandwidth of 672\,MHz. 

Across the full 1167\,hr observing campaign, we detected 97 bursts from FRB\,20240114A over a frequency range extending from approximately 900\,MHz to 5\,GHz (see Figure \ref{fig:frequency_vs_MJD}). Table~\ref{tab:obs_table} summarizes the observing time as a function of MJD and frequency tuning, together with the number of detected bursts and the corresponding burst rate.

\begin{figure*}
    \centering
    \includegraphics[width=0.73\linewidth]{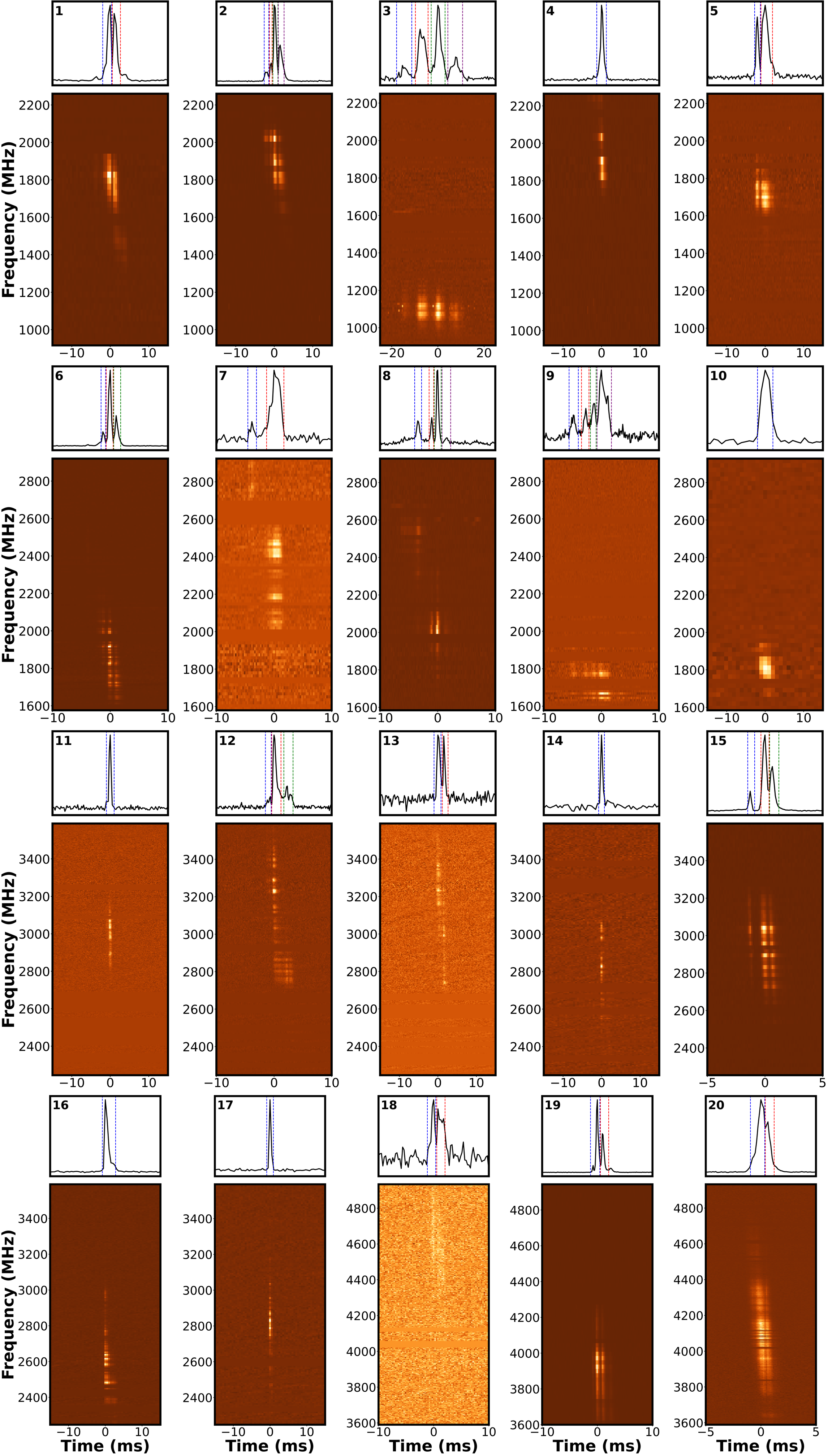}
    \caption{Dynamic Spectra of 20 of the detected FRBs. The vertical dotted lines represent the on-pulse regions of the bursts as well as the sub-bursts, which are used for the SNR calculations described in Section \ref{ssec:SNR_flux_fluence}. The dynamic spectrum plots for all 97 bursts can be found on \href{https://zenodo.org/uploads/19429415}{zenodo.org/uploads/19429415}. }
    \label{fig:dynamic_spectrum}
\end{figure*}

The dynamic spectra of the bursts, incoherently dedispersed at a DM of 528.1 $\mathrm{pc\ cm^{-3}}$, after calculating the structure-optimized DM using \texttt{DM\_phase} \citep{michilli_DM_phase} on the bright bursts, are shown in Fig. \ref{fig:dynamic_spectrum}. This value was calculated by taking the median of the list of structure-optimized DMs across the LO tunings and several bright bursts. The final value we got is 528.1 $\pm$ 0.2 $\mathrm{pc\ cm^{-3}}$ as the median value for the structure-optimized DM, which was used in all the further analysis of the bursts.

\section{Analysis}
\label{sec:burst_properties}
\subsection{File Preparation}
\label{ssec:File_preparation}
After identifying a burst in a 30-min filterbank file for a given LO tuning, to extract fine structures in the burst, we returned to the corresponding raw, pre-decimation data and extracted the native 32-bit filterbank files for both simultaneous LO tunings from that scan. For each burst, the two simultaneous LO filterbank files were then spliced together to produce a single 32-bit filterbank file with a total bandwidth of 1344\,MHz. We extracted a 5-s segment centered on the burst arrival time reported by \texttt{SPANDAK}. This procedure was applied to all 97 detected bursts.

The burst segments were then processed with \texttt{DSPSR} \citep{dspsr_cite} to generate multiple \textit{PSRFITS} files, each spanning 130\,ms. The files were dedispersed at the structure-optimized DM of 528.1\,pc\,cm$^{-3}$. For each burst, we selected the \textit{PSRFITS} file containing the burst and centered it for further analysis. The resulting files were manually cleaned using \texttt{PAZI} from \texttt{PSRCHIVE} to remove any remaining spurious RFI. After manual zapping, the \textit{PSRFITS} files were averaged at a time and frequency resolution that was manually selected to maximize SNR while preserving as much spectrotemporal structure as possible. Finally, the cleaned and binned files were converted to \texttt{npz} format, following the metadata requirements of \texttt{FRBGUI} \citep{Chamma_2023_FRBGUI}, for the spectro-temporal analysis described in the Section \ref{ssec:FRBGUI}. 

\subsection{Calculating Burst S/N, Flux Density, and Fluence}
\label{ssec:SNR_flux_fluence}

Using the calibration scans described in Section~\ref{ssec:Observations}, we derived the flux densities and fluences of the detected bursts from FRB\,20240114A. The analysis first required estimating the system-equivalent flux density (SEFD) for each observing session. We observed standard flux calibrators whose spectral flux-density models are given by \citet{Perley_Butler_2017}. From these models, we evaluated the expected calibrator flux density, $S_{\mathrm{exp}}$, at the central frequency of each LO tuning. For each antenna--polarization pair $(x,y)$, the SEFD was estimated as
\begin{equation}
    \mathrm{SEFD}_{x,y} =
    \left(|G_{\mathrm{cal}}|^{2} S_{\mathrm{exp}}\right)^{-1},
\end{equation}
where $G_{\mathrm{cal}}$ is the complex antenna-based gain solution obtained from the corresponding calibration scan using the \texttt{gaincal} task in the Common Astronomy Software Applications package \citep[CASA;][]{CASA_cite_2022}. We use the squared gain amplitude, $|G_{\mathrm{cal}}|^{2}$, since the CASA gain solutions are complex. In this convention, $|G_{\mathrm{cal}}|^{2} S_{\mathrm{exp}}$ represents the calibrated response of a given antenna--polarization signal chain to the flux calibrator, and its inverse gives the corresponding SEFD estimate.

For each LO tuning, we then estimated the beamformed SEFD by averaging over all available antenna--polarization pairs and scaling by the number of antennas used in the coherent beamformer,
\begin{equation}
    \mathrm{SEFD}_{\mathrm{BF}} =
    \frac{1}{N_{\mathrm{ant}}}
    \left(
    \frac{1}{N_{x,y}}
    \sum_{x,y} \mathrm{SEFD}_{x,y}
    \right),
\end{equation}
where $N_{x,y}$ is the number of antenna--polarization pairs and $N_{\mathrm{ant}}$ is the number of antennas included in the beamformer. For the 20-antenna and 28-antenna configurations, these correspond to $N_{x,y}=40$ and $56$, respectively. The SEFDs were therefore calculated separately for the two array configurations.

This procedure yielded a polarization-averaged beamformer SEFD for each LO tuning in each observing scan. Since two LO tunings were recorded simultaneously, we combined their SEFDs to obtain an effective SEFD for the full 1344\,MHz bandwidth. For two LO tunings with SEFDs $a$ and $b$, the combined SEFD was calculated as
\begin{equation}
    \mathrm{SEFD}_{\mathrm{comb}} =
    \frac{a\,b}{\sqrt{a^{2}+b^{2}}}.
\end{equation}
This combined SEFD was then used to estimate the flux densities and fluences of the bursts detected in that scan. We estimated the minimum detectable flux density using the radiometer equation,  

\begin{equation}
    S_{\nu,\mathrm{min}} =
    \frac{(\mathrm{S/N})_{\mathrm{min}}\,\mathrm{SEFD}}
    {\sqrt{n_{\mathrm{pol}}\,\tau\,\Delta\nu}},
\end{equation}
where $n_{\mathrm{pol}}=2$ is the number of polarizations, $\tau$ is the assumed burst duration, and $\Delta\nu$ is the effective bandwidth. The corresponding minimum fluence threshold for each observing configuration is listed in Table~\ref{tab:obs_table}.

For each burst or sub-burst, the signal-to-noise ratio was measured from the normalized time profile as
\begin{equation}
    \mathrm{S/N} =
    \frac{\sum p_i}{\sqrt{N}},
\end{equation}
where $p_i$ is the normalized profile amplitude in the on-pulse region and $N$ is the number of time bins included in the burst or sub-burst window. The on-pulse regions used for these measurements are marked by vertical dotted lines in Figure~\ref{fig:dynamic_spectrum}. The same windows were used to calculate the corresponding flux densities and fluences.

The flux density was then calculated as
\begin{equation}
    S_{\mathrm{Jy}} =
    \frac{\mathrm{SEFD}\,(\mathrm{S/N})}
    {\sqrt{B\,N_{\mathrm{pol}}\,N\,t_{\mathrm{samp}}}},
\end{equation}
where $B$ is the sub-burst bandwidth, $N_{\mathrm{pol}}=2$ is the number of polarizations, $N$ is the number of time bins across the burst or sub-burst, and $t_{\mathrm{samp}}$ is the sampling interval. Fluences were obtained by multiplying $S_{\mathrm{Jy}}$ by the corresponding burst or sub-burst duration in milliseconds. The measured S/N, flux density, and fluence values for a representative sample of 20 bursts, including their identified sub-bursts, are reported in Table~\ref{tab:frb_properties}.

\subsection{Spectro-temporal Properties with FRBGUI}
\label{ssec:FRBGUI}
We measured the spectro-temporal properties of the 97 detected bursts from FRB\,20240114A, including their durations, bandwidths, center frequencies, and drift rates, using the interactive analysis package \texttt{FRBGUI} \citep{Chamma_2023_FRBGUI}. The dedispersed and cropped dynamic spectra, converted to \texttt{numpy} arrays as described in Section~\ref{ssec:File_preparation}, were used as inputs to \texttt{FRBGUI}. All burst properties reported here were measured after dedispersing the data at the structure-optimized DM of 528.1\,pc\,cm$^{-3}$. Although \texttt{FRBGUI} allows the user to evaluate burst properties over a grid of trial DMs, we adopted this single DM consistently for all bursts and sub-bursts to avoid introducing burst-to-burst differences caused by independent DM choices.

Several optional processing steps were applied manually, when needed, to improve the burst S/N and reduce the impact of residual RFI. First, a background region, typically selected from the first few milliseconds preceding the burst, was subtracted from the full dynamic spectrum. Second, for bursts that occupied only part of the observing band, frequency ranges without detectable burst emission were masked. In some cases, this required excluding several hundred MHz of bandwidth and significantly improved the measured S/N. Third, the dynamic spectra were downsampled in time and/or frequency for weaker bursts. The downsampling factors were chosen to evenly divide the data dimensions while preserving sufficient resolution to measure temporal substructure and frequency drift.

Burst components were then identified manually using the interactive \texttt{FRBGUI} interface. We marked all visibly distinct sub-bursts within the available time and frequency resolution. Of the 97 bursts in our sample, 33 showed multiple components, yielding a total of 162 components, including both full bursts and sub-bursts, analyzed with \texttt{FRBGUI}. A small number of bursts contained low-S/N sub-components that could not be reliably separated; these events were therefore analyzed as single components.

For quantitative characterization, \texttt{FRBGUI} fits a six-parameter two-dimensional Gaussian model to the two-dimensional autocorrelation function of the dynamic spectrum. We first applied this procedure to the full burst window and then repeated it for the time-restricted sub-arrays corresponding to each identified sub-burst. Fitting the autocorrelation function improves the effective S/N and reduces the influence of scintillation and residual RFI on the measured parameters. At the adopted DM of 528.1\,pc\,cm$^{-3}$, the fitting routine returned the burst bandwidth, center frequency, duration, and slope for individual sub-bursts, as well as the corresponding properties for the full burst, including the drift rate and associated uncertainties using methods discussed in \cite{Chamma_2023_FRBGUI}. 

In cases where the fitting procedure did not converge to physically meaningful values, the fits were inspected manually within the \texttt{FRBGUI} interface. For eight bursts, reliable fitted parameters could not be obtained because of low S/N or strong scattering/scintillation structure. For these bursts, we report visually estimated values (lower limits) and mark the corresponding uncertainties as \texttt{NA}.

The burst times of arrival are reported as topocentric MJDs. The spectro-temporal properties for a representative subset of bright bursts and their sub-bursts are listed in Table~\ref{tab:frb_properties}. The reported durations and bandwidths correspond to 3-$\sigma$ widths converted from the full-width at half-maximum values returned by \texttt{FRBGUI}. The uncertainties listed in parentheses were converted in the same way. When the formal fitting uncertainty was smaller than the instrumental resolution, we adopted minimum uncertainties corresponding to the time and frequency resolutions, equivalent to 0.15\,ms in duration and 1\,MHz in center frequency or bandwidth after conversion to 3-$\sigma$ values.

We also estimated the isotropic-equivalent burst energy for each burst and sub-burst following Equation~3 of \citet{Aggarwal_2021},
\begin{equation}
E = 4\pi
\left( \frac{D_L}{\mathrm{cm}} \right)^2
\left( \frac{F}{\mathrm{Jy\,s}} \right)
\left( \frac{\Delta \nu_{\mathrm{occ}}}{\mathrm{Hz}} \right)
\times 10^{-23}\ \mathrm{erg},
\end{equation}
where $D_L=630.72$\,Mpc is the luminosity distance of FRB\,20240114A, corresponding to a redshift of $z=0.13$ \citep{redshift_ATel} and the cosmological parameters from the Planck Collaboration \citep{Planck_2020}. Here, $F$ is the measured fluence and $\Delta \nu_{\mathrm{occ}}$ is the occupied bandwidth of the burst. Both quantities were derived from the \texttt{FRBGUI}-based measurements. The resulting isotropic-equivalent energies are reported in Table~\ref{tab:frb_properties}.

For multi-component bursts, we report drift rates only for the combined component structure. Negative drift-rate values indicate the characteristic downward frequency drift of the bursts. 


\setlength{\tabcolsep}{2pt}
\renewcommand{\arraystretch}{1.15}

\begin{longtable*}{cccccccccc}

\caption{Spectro-temporal properties of 20 bursts of FRB 20240114A detected by the Allen Telescope Array, as per the indexing order from Figure ~\ref{fig:dynamic_spectrum}. The first entry reports the complete burst properties, while all subsequent entries for that burst correspond to the sub-burst properties. The full table of spectro-temporal properties for all 97 bursts is available as a CSV file at \href{https://zenodo.org/uploads/19429571}{zenodo.org/uploads/19429571}.}
\\
\hline
Index & Burst MJD & SNR & Duration & Center Freq. & Bandwidth & Flux & Fluence & Drift Rate & Energy \\
& (Topocentric) &  & (ms) & (MHz) & (MHz) & (Jy) & (Jy-ms) & (MHz ms$^{-1}$) & ($10^{39}$ ergs) \\
\hline
\endfirsthead

\multicolumn{10}{c}%
{{\bfseries Table \thetable\ (continued)}}\\
\hline
Index & Burst MJD & SNR & Duration & Cent Freq. & Bandwidth & Flux & Fluence & Drift Rate & Energy \\
 & (Topocentric) & & (ms) & (MHz) & (MHz) & (Jy) & (Jy ms) & (MHz ms$^{-1}$) & ($10^{39}$ ergs) \\
\hline
\endhead

1 & 60396.75388296296 & 269.42 & $3.5(1.0)$ & $1754(42)$ & $840(42)$ & 31.90 & 110.10 & $-174.18 \pm 2.43$ & 44.00 \\
 &   & 192.99 & $1.3(5)$ & $1790(24)$ & $607(24)$ & 43.25 & 57.68 & -- & 16.70 \\
 &   & 188.03 & $1.3(5)$ & $1714(24)$ & $829(24)$ & 36.17 & 47.97 & -- & 18.90 \\
2 & 60401.594488981485 & 496.62 & $3.7(1.0)$ & $1866(42)$ & $1075(42)$ & 49.62 & 183.77 & $-156.89 \pm 1.35$ & 94.00 \\
 &   & 16.7 & $1.2(3)$ & $1988(16)$ & $618(16)$ & 3.81 & 4.71 & -- & 1.38 \\
 &   & 6.41 & $0.5(3)$ & $1984(16)$ & $725(16)$ & 2.09 & 1.08 & -- & 0.37 \\
 &   & 579.02 & $1.0(3)$ & $1898(16)$ & $860(16)$ & 126.94 & 122.13 & -- & 50.00 \\
 &   & 347.29 & $1.7(3)$ & $1802(16)$ & $797(16)$ & 59.98 & 100.23 & -- & 38.00 \\
3 & 60424.65145706019 & 96.8 & $24.9(4.1)$ & $1063(42)$ & $435(42)$ & 6.23 & 155.14 & $-29.19 \pm 2.22$ & 32.10 \\
 &   & 12.25 & $5.1(5)$ & $1101(12)$ & $373(12)$ & 1.87 & 9.64 & -- & 1.71 \\
 &   & 93.17 & $5.1(5)$ & $1084(12)$ & $382(12)$ & 14.18 & 71.93 & -- & 13.10 \\
 &   & 62.84 & $4.6(5)$ & $1077(12)$ & $368(12)$ & 10.27 & 46.99 & -- & 8.22 \\
 &   & 28.1 & $5.9(5)$ & $1071(12)$ & $379(12)$ & 3.98 & 23.48 & -- & 4.24 \\
4 & 60428.67938119212 & 138.72 & $1.7(1)$ & $1862(42)$ & $806(42)$ & 25.04 & 41.84 & -- & 16.10 \\
5 & 60435.60110179398 & 146.35 & $4.5(2.0)$ & $1684(16)$ & $528(16)$ & 20.46 & 91.27 & $-293.41 \pm 10.23$ & 22.90 \\
 &   & 58.24 & $1.2(1)$ & $1709(1)$ & $491(1)$ & 16.22 & 19.57 & -- & 4.58 \\
 &   & 136.73 & $2.8(1)$ & $1697(1)$ & $460(1)$ & 25.66 & 72.91 & -- & 16.00 \\
6 & 60436.56506990441 & 410.69 & $2.3(1)$ & $1844(2)$ & $1052(2)$ & 61.53 & 144.27 & $-239.83 \pm 0.36$ & 72.20 \\
 &   & 69.04 & $0.9(1)$ & $1974(2)$ & $832(2)$ & 18.97 & 16.72 & -- & 6.62 \\
 &   & 544.87 & $0.7(1)$ & $1862(2)$ & $890(2)$ & 168.31 & 109.68 & -- & 46.50 \\
 &   & 145.79 & $0.9(1)$ & $1729(2)$ & $699(2)$ & 42.26 & 39.85 & -- & 13.30 \\
7 & 60440.57816840278 & 46.47 & $2.6(5)$ & $2227(4)$ & $1636(5)$ & 5.32 & 13.69 & $-934.30 \pm 13.30$ & 10.70 \\
 &   & 7.99 & $1.2(5)$ & $2269(87)$ & $1135(5)$ & 1.59 & 1.95 & -- & 1.06 \\
 &   & 51.31 & $2.5(5)$ & $2314(4)$ & $1615(4)$ & 6.06 & 14.83 & -- & 11.40 \\
8 & 60452.55498287037 & 48.68 & $4.0(3)$ & $2090(4)$ & $2126(7)$ & 3.71 & 14.85 & $-147.29 \pm 0.43$ & 15.00 \\
 &   & 30.71 & $3.4(3)$ & $2466(4)$ & $759(4)$ & 4.26 & 14.41 & -- & 5.21 \\
 &   & 20.56 & $0.6(3)$ & $2017(4)$ & $419(4)$ & 9.31 & 5.36 & -- & 1.07 \\
 &   & 62.18 & $0.6(3)$ & $2004(4)$ & $471(4)$ & 27.12 & 14.98 & -- & 3.36 \\
 &   & 5.17 & $2.2(3)$ & $2018(9)$ & $794(4)$ & 0.87 & 1.91 & -- & 0.72 \\
9 & 60475.53778605324 & 51.96 & $6.9(5)$ & $1700(6)$ & $662(6)$ & 4.61 & 31.88 & $-76.61 \pm 0.86$ & 10.00 \\
 &   & 13.04 & $1.9(5)$ & $1759(6)$ & $381(6)$ & 2.89 & 5.58 & -- & 1.01 \\
 &   & 15.16 & $1.0(5)$ & $1748(6)$ & $545(6)$ & 3.88 & 3.92 & -- & 1.02 \\
 &   & 20.82 & $5.2(5)$ & $1753(6)$ & $506(6)$ & 2.44 & 12.65 & -- & 3.05 \\
 &   & 54.88 & $2.7(5)$ & $1682(6)$ & $665(6)$ & 7.77 & 21.03 & -- & 6.65 \\
10 & 60478.63422543982 & 51.02 & $3.6(1.0)$ & $1815(24)$ & $482(24)$ & 6.99 & 24.86 & -- & 5.71 \\
11 & 60499.517409722226 & 34.49 & $0.7(5)$ & $2920(11)$ & $1371(7)$ & 6.98 & 5.03 & -- & 3.28 \\
12 & 60493.3478609375 & 56.76 & $0.7(3)$ & $3001(2)$ & $725(2)$ & 16.81 & 11.40 & -- & 3.93 \\
13 & 60495.440898067136 & 87.27 & $2.6(3)$ & $3019(6)$ & $2322(6)$ & 7.14 & 18.35 & $-296.84 \pm 0.66$ & 20.30 \\
 &   & 9.47 & $0.7(3)$ & $3231(14)$ & $1154(6)$ & 2.08 & 1.49 & -- & 0.82 \\
 &   & 104.26 & $1.1(3)$ & $3146(6)$ & $1982(6)$ & 13.92 & 15.73 & -- & 14.80 \\
 &   & 34.81 & $1.7(3)$ & $2808(6)$ & $469(6)$ & 7.81 & 13.20 & -- & 2.95 \\
14 & 60497.35840775463 & 28.21 & $1.9(5)$ & $3125(4)$ & $1968(6)$ & 3.03 & 5.67 & $-378.32 \pm 1.36$ & 5.31 \\
 &   & 23.65 & $1.0(5)$ & $3225(5)$ & $1120(4)$ & 4.68 & 4.53 & -- & 2.42 \\
 &   & 15.92 & $0.8(5)$ & $3019(15)$ & $1572(4)$ & 2.94 & 2.33 & -- & 1.74 \\
15 & 60499.524800590276 & 437.22 & $2(1)$ & $2938(42)$ & $1294(42)$ & 54.75 & 109.17 & $-1493.41 \pm 81.37$ & 67.20 \\
 &   & 43.77 & $0.3(3)$ & $3017(4)$ & $905(4)$ & 16.90 & 5.07 & -- & 2.18 \\
 &   & 440.5 & $0.4(3)$ & $2952(4)$ & $1096(4)$ & 129.11 & 55.50 & -- & 28.90 \\
 &   & 241.92 & $0.5(3)$ & $2913(4)$ & $1176(4)$ & 66.37 & 30.33 & -- & 17.00 \\
16 & 60501.516398831016 & 182.54 & $1.5(5)$ & $2557(6)$ & $1040(6)$ & 29.89 & 44.17 & -- & 21.90 \\
17 & 60503.585854282406 & 64.05 & $0.6(5)$ & $2825(8)$ & $844(8)$ & 18.89 & 10.76 & -- & 4.32 \\
18 & 60523.42831423611 & 20.83 & $3.5(5)$ & $4337(15)$ & $1834(14)$ & 2.16 & 7.57 & $-708.78 \pm 23.78$ & 6.61 \\
 &   & 13.2 & $1.3(5)$ & $4438(49)$ & $1513(4)$ & 2.46 & 3.24 & -- & 2.33 \\
 &   & 16.48 & $1.1(5)$ & $4320(55)$ & $1810(4)$ & 3.12 & 3.33 & -- & 2.87 \\
19 & 60538.19848472222 & 927.67 & $2.2(1.0)$ & $3896(24)$ & $1070(24)$ & 168.56 & 374.17 & $-1021.63 \pm 34.88$ & 191.00 \\
 &   & 818.16 & $0.6(3)$ & $3918(2)$ & $840(2)$ & 317.71 & 196.70 & -- & 78.60 \\
 &   & 517.96 & $0.8(3)$ & $3882(2)$ & $985(2)$ & 161.46 & 132.27 & -- & 62.00 \\
20 & 60546.24161284722 & 876.88 & $1.6(1)$ & $4086(1)$ & $1585(1)$ & 131.68 & 203.92 & $-893.06 \pm 0.28$ & 154.00 \\
 &   & 758.98 & $0.9(1)$ & $4119(1)$ & $1515(1)$ & 149.10 & 141.17 & -- & 102.00 \\
 &   & 455.75 & $0.6(1)$ & $4016(1)$ & $1480(1)$ & 112.31 & 69.18 & -- & 48.70 \\
\hline
\label{tab:frb_properties}
\end{longtable*}

\begin{figure*}
    \centering
    \includegraphics[width=1.0\linewidth]{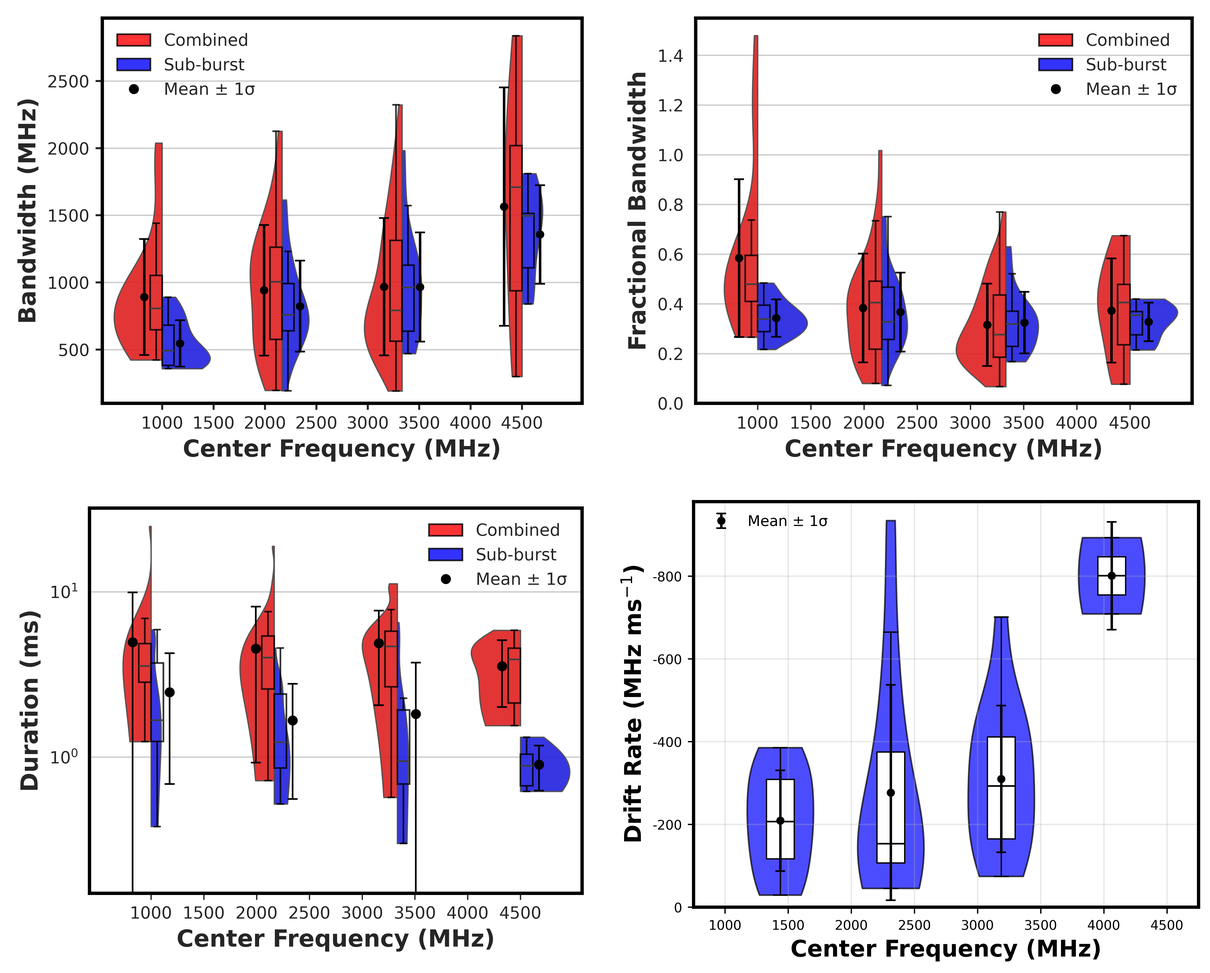}
    \caption{Distributions of spectro-temporal burst properties as a function of center frequency for FRB\,20240114A. The X-axis is binned by a factor of 4 to account for the 4 unique LO tunings in which we detected all the bursts. The Split violin plots compare measurements from the full burst envelopes with those from the individual sub-burst components. The left half of each violin shows the full-burst measurements, while the right half shows the sub-burst measurements. Embedded box plots indicate the median and interquartile range, and black points with vertical error bars mark the mean and standard deviation of each distribution. The panels show the bandwidth, fractional bandwidth, duration, and drift rate.}
    \label{fig:subburst_properties_analysis_binned}
\end{figure*}

Figure~\ref{fig:subburst_properties_analysis_binned} shows the distributions of the measured spectro-temporal properties as a function of burst center frequency. However, note that the X-axis in the plot is binned by a factor of 4 to account for the 4 unique LO tunings in which we detected all the bursts.  We compare the properties measured from the full burst envelopes with those measured from the individual sub-burst components. The parameters shown are bandwidth, fractional bandwidth, duration, and drift rate. For each frequency bin, the distributions are shown using split violin plots with embedded box plots, allowing the full distribution of values to be compared between the full-burst and sub-burst measurements.

The width of each violin represents the relative density of measurements at a given value, while the embedded box plot shows the interquartile range. The central line marks the median, and the whiskers indicate the spread of the distribution. Black points and vertical error bars show the mean and standard deviation for each distribution. This representation highlights both the typical values and the intrinsic scatter of the burst properties across the observed frequency range.

\section{Discussion}
\label{sec:analysis}
\subsection{Frequency dependent burst activity}
FRB\,20240114A exhibits pronounced frequency-dependent activity, a property well established in repeating FRBs such as the periodic source FRB\,20180916B \citep{pleunis_lofar_2021} and now mapped for this source across more than a decade in frequency, from 327\,MHz \citep{Westerbork_Atel} and the GMRT bands \citep{Ajay_GMRT_ApJ,Ujjwal_GMRT_paper} to the 6\,GHz Effelsberg detections \citep{limaye2026}. Across our 1167\,hr ATA campaign we detected 97 bursts from $\sim$900\,MHz to $\sim$5\,GHz, with a burst rate that varies strongly in both frequency and time, peaking at $0.25$\,hr$^{-1}$ in the 2244--3588\,MHz band during MJD\,60480--60510 and demonstrating that the source has been extremely active well above 2\,GHz (Table~\ref{tab:obs_table}). This burst storm is highly unusual and, to the best of our knowledge, has not been reported before for this FRB. Despite $\sim$305\,hr spanning 4932--7620\,MHz we detect no bursts in these highest bands, confirming that the emission, while broadband, remains band-limited in any single epoch, consistent with the strong high-frequency suppression seen at 8.6\,GHz by \citet{shangai_longterm_paper} and the 6\,GHz envelope of \citet{limaye2026}. 

\begin{figure*}
    \centering
    \includegraphics[width=1.0\linewidth]{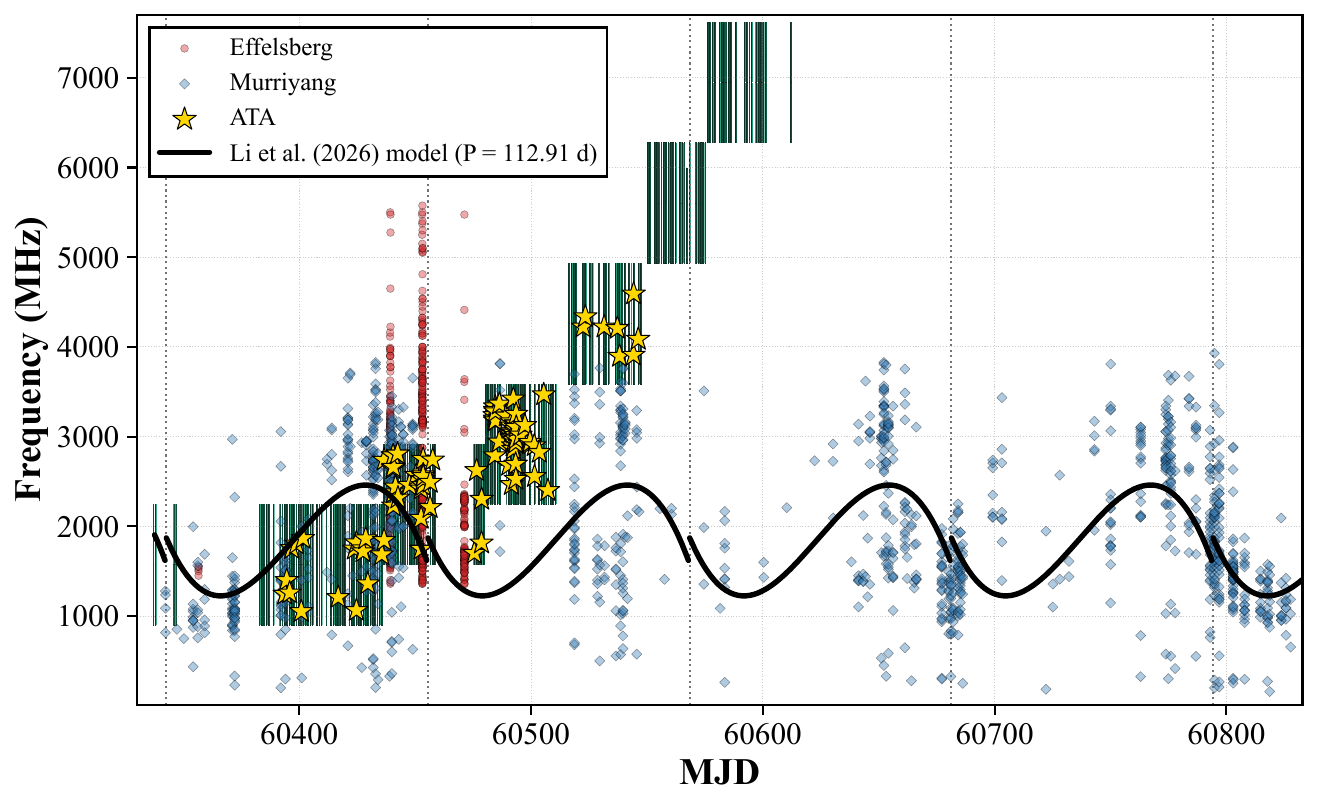}
        \caption{Multi-telescope burst center frequency versus MJD, with the $P=112.91$\,day periodic frequency-modulation model of \citet{Li2026_freqperiod} overlaid (solid black curve; dotted lines mark cycle boundaries). ATA (gold stars), Murriyang (blue diamonds; \cite{Uttarkar2026}), and Effelsberg (red circles; \cite{limaye2026}) bursts are shown as in Figure~\ref{fig:frequency_vs_MJD}, with ATA LO coverage indicated by the blocks of narrow vertical lines. The Murriyang sample broadly follows the model, but the ATA high-frequency burst storm near MJD\,60480--60546 lies above the curve at a phase where low-frequency emission is predicted, contradicting a strictly phase-coherent modulation by \cite{Li2026_freqperiod}.}
    \label{fig:li_model_overlay}
\end{figure*}

Two periodicities have recently been claimed for the source. \citet{FAST_periodicity_paper2} report a $143.4\pm7.2$\,day periodicity in the burst \emph{activity} from FAST, whereas \citet{Li2026_freqperiod}, analysing the Murriyang sample of \citet{Uttarkar2026}, find a ($>6\sigma$) $\sim$112.9\, day periodicity in the \emph{central emission frequency}, a low-to-high sawtooth drift within each cycle, that is decoupled from the burst rate, since neither \citet{Uttarkar2026} nor \citet{Li2026_freqperiod} detect any periodicity in the arrival times themselves. The ATA campaign cannot test these long periods directly: our detections span only $\sim$153\, days (shorter than 1.5 cycles), and our deliberately upward-stepped tuning schedule makes the ATA central frequency correlate with epoch by construction (Pearson $r=0.86$), in the same sense as the claimed drift but degenerate with it. We therefore use the ATA data not to fit a period but to test whether our detections are \emph{consistent} with the proposed phase coherence, by overlaying the \citet{Li2026_freqperiod} model on the multi-telescope burst data (Figure~\ref{fig:li_model_overlay}).

As shown in Figure~\ref{fig:li_model_overlay}, the central-frequency modulation reported by \cite{Li2026_freqperiod} is not fully consistent with our ATA detections. Near MJD\,60500, corresponding to phase $\sim$0.3--0.48 of the proposed 112.9-day cycle, \cite{Li2026_freqperiod} model predicts predominantly low-frequency emission, whereas the ATA detected its strongest high-frequency burst storm, with 45 bursts centered at 2.4--3.5\,GHz. This activity was largely missed by other campaigns: Murriyang detected only seven bursts in the same interval because of observing gaps, and Effelsberg was not observing. Although our limited frequency coverage cannot determine whether simultaneous low-frequency emission was present, these detections show that a substantial S-band activity episode was not captured by the datasets used to infer the proposed modulation by \cite{Li2026_freqperiod}. Its omission may have contributed to the apparent 112.9-day frequency periodicity, indicating that incomplete frequency-time coverage must be accounted for before the proposed $\sim$112.9-, and $\sim$143.4-day modulations can be considered established. Continued wideband, high-cadence monitoring is required to test whether the frequency modulation persists or breaks down when the full spectral activity of FRB\,20240114A is sampled.

\subsection{Bandwidth Evolution with Frequency}
The bandwidth distributions in the top-left panel of Figure~\ref{fig:subburst_properties_analysis_binned} show a clear increase of absolute burst bandwidth with center frequency, traced by the rising mean values (black dots). The broader bandwidths of the combined bursts arise in part from the superposition of multiple sub-bursts, each occupying a narrower spectral range. A growth of absolute bandwidth toward higher frequencies is expected if the emission occupies a roughly fixed fractional bandwidth, and is in line with the $\sim$1\,GHz-wide spectral features seen for FRB\,20121102A in the 4--8\,GHz band by \citet{gajjar_highest_2018}.

The fractional bandwidth, $\Delta \nu / \nu_c$, where $\Delta \nu$ is the bandwidth and $\nu_c$ is the center frequency, provides a normalized measure of spectral occupancy that is only weakly dependent on frequency in our sample (top-right panel of Figure~\ref{fig:subburst_properties_analysis_binned}). Its approximate scale-invariance agrees with the broadband studies of FRB\,20240114A by \citet{limaye2026} and \citet{energydist_paper}, and with FRB\,20121102A \citep{gajjar_highest_2018}, all of which find no strong frequency evolution of the fractional bandwidth. The \emph{absolute} value we measure, however, is larger: our 3-$\sigma$ fractional bandwidths have a median of $\sim$35\% (combined bursts and sub-bursts), whereas \citet{limaye2026} and \citet{energydist_paper} report $\sim$10\% for this source and repeaters more generally occupy $\sim$10--30\% \citep{pleunis_lofar_2021}. Part of this offset is definitional, we quote 3-$\sigma$ Gaussian widths, which substantially exceed full-width-half-maximum or visually-defined extents (the FWHM-based median of our sample is $\sim$14\%, much closer to the literature). The remaining excess is most plausibly a sensitivity effect: the ATA is markedly less sensitive than FAST or the 100-m Effelsberg dish (Table~\ref{tab:obs_table}), so our sample is restricted to the brightest, highest-fluence bursts, which are preferentially the more broadband events. This selection biases our measured fractional-bandwidth distribution upward relative to the deeper, lower-fluence samples of \citet{limaye2026} and \citet{energydist_paper}. As expected if the combined bursts are superpositions of narrower components, the sub-burst fractional-bandwidth distributions are narrower than those of the combined bursts, supporting the interpretation that sub-bursts represent the fundamental emission units.

\subsection{Burst width with Frequency}
The sub-burst duration decreases with increasing observing frequency, as seen in the bottom-left panel of Figure~\ref{fig:subburst_properties_analysis_binned}: sub-bursts at lower frequencies have longer durations and a wider spread, while those at higher frequencies are shorter. This inverse scaling agrees with other studies of FRB\,20240114A and of repeaters generally. \citet{limaye2026} fit a power law $W\propto\nu^{\alpha}$ to their uniform 1.3--6\,GHz Effelsberg widths and obtain $\alpha\simeq-0.5$ to $-1.0$, and \citet{shangai_longterm_paper} measure a median width of only $\sim$3\,ms at 2.25\,GHz, narrower than the widths reported at L-band and below. The same behavior was found for FRB\,20121102A by \citet{gajjar_highest_2018}, whose 4--8\,GHz bursts had relatively narrow average widths of $\sim$0.64\,ms, considerably shorter than those measured at $\lesssim$3\,GHz. The narrowing of sub-bursts toward higher frequencies is qualitatively consistent with the reduced influence of multi-path scattering, which broadens profiles more strongly at lower frequencies \citep{extrinsic_model}, although intrinsic spectral evolution such as radius-to-frequency mapping of the emission region cannot be excluded \citep{intrinsic_model}.

\subsection{Drift Rate with Frequency}\label{subsec:driftvfreq}
The drift-rate distributions show an increase in magnitude with observing frequency, as seen in the bottom-right panel of Figure~\ref{fig:subburst_properties_analysis_binned}. The bursts exhibit predominantly negative drift rates, corresponding to downward drift in frequency with time, the ``sad-trombone'' behavior commonly observed in repeating FRBs. The distribution is especially broad near 2.3\,GHz, indicating substantial burst-to-burst diversity in the measured drift rates at this frequency. The increase of drift-rate magnitude with frequency is consistent with the FAST burst-cluster analysis of \citet{FAST_drifting_paper3}, who found that the drift rate of single-component clusters increases with peak frequency. This trend may be interpreted in the context of intrinsic emission models, including radius-to-frequency mapping in neutron-star magnetospheres, in which the emission frequency changes with altitude in the source region \citep{intrinsic_model}. Related spectro-temporal correlations, including the dependence of sub-burst slope on duration and observing frequency, have also been discussed in the framework of the triggered relativistic dynamical model \citep[e.g.,][]{Rajabi2020,Chamma2021,Chamma2025}. Propagation effects, including plasma lensing in a magnetized circumburst medium, may also contribute to the observed drift-rate behavior \citep{plasma_lensing,extrinsic_model}.

\citet{FAST_drifting_paper3} further report that downward-drifting bursts are the more energetic events, whereas upward drifting is largely confined to lower-energy, narrower, and fainter bursts. Our sample is consistent with this picture: it is dominated by downward-drifting, more energetic bursts, and we find no evidence for a population of bright upward drifters. 

\subsection{Isotropic Burst Energy with Frequency}
\begin{figure*}
    \centering
    \includegraphics[width=0.8\linewidth]{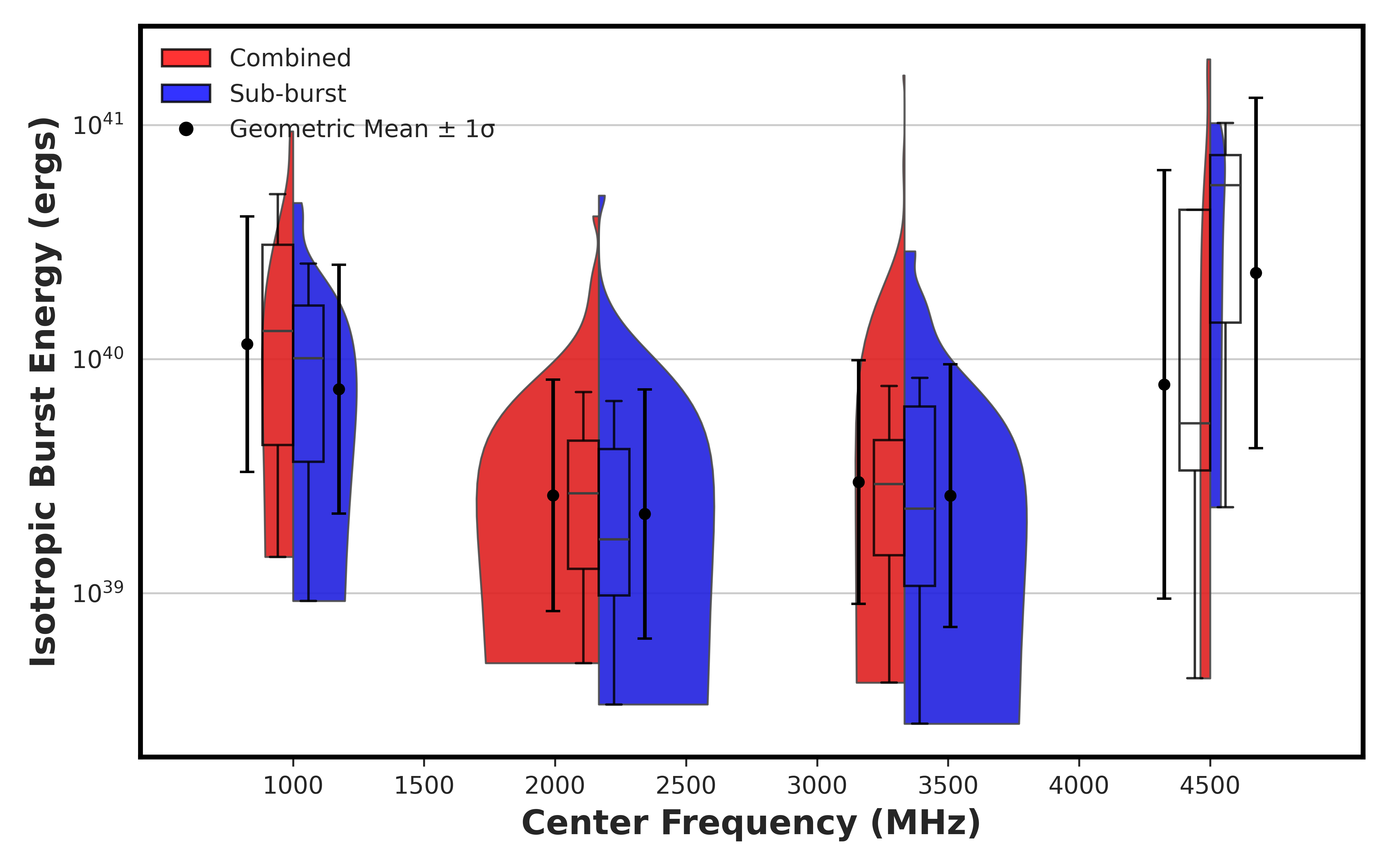}
    \caption{Isotropic-equivalent burst energy as a function of center frequency for FRB\,20240114A. Red and blue violin plots show the distributions for combined bursts and individual sub-bursts, respectively. Embedded box plots indicate the median and interquartile range, while black points and vertical bars mark the geometric mean and standard deviation. The distributions span several orders of magnitude, indicating a broad energy range at each observed frequency.}
    \label{fig:energy_plot}
\end{figure*}

Figure~\ref{fig:energy_plot} shows the distribution of isotropic-equivalent burst energies as a function of center frequency for the combined bursts and sub-bursts. The energies span roughly $10^{37}$--$10^{41}$\,erg, with broad, non-Gaussian distributions in each frequency bin. Combined bursts generally reach higher energies than their individual sub-bursts, as expected when multiple components contribute to the total fluence. The sub-bursts occupy a narrower bandwidth and lower energy ranges, supporting the interpretation that they represent more elementary emission components.

We do not find a simple monotonic dependence of isotropic-equivalent energy on observing frequency. The highest-frequency ATA bursts, however, occupy the upper end of the energy distribution and show substantial scatter, demonstrating that energetic bursts from FRB\,20240114A are not confined to L band. This is important for the source energy budget. FAST observations at L band detected more than $10^4$ bursts and argued that, after accounting for duty cycle, beaming, and radio efficiency, the cumulative emitted energy approaches a large fraction of the magnetic-energy reservoir of a typical magnetar \citep{energydist_paper}. Similarly, the HyperFlash campaign showed that rare, high-fluence bursts can dominate the cumulative energy release, with the brightest events contributing disproportionately to the total energy budget \citep{HyperFlash2026}. Our ATA detections extend this picture to higher radio frequencies: although our sample is much smaller and less sensitive to the low-energy burst population, it shows that S-band and higher-frequency bursts can contribute non-negligibly to the total emitted radio energy.

The apparent frequency dependence in Figure~\ref{fig:energy_plot} should therefore be interpreted cautiously. Because the ATA frequency coverage was stepped with epoch, the energy distribution reflects both intrinsic burst-energy variability and the evolving activity state of the source. Nevertheless, the presence of high-energy bursts above 3\,GHz implies that energy-budget estimates based only on L-band activity are incomplete. A full accounting of the radio energy released by FRB\,20240114A requires wideband, high-cadence monitoring capable of capturing both the numerous lower-energy bursts detected by sensitive facilities such as FAST and the rarer, high-energy bursts emphasized by HyperFlash and detected here at higher frequencies with the ATA. 

\subsection{Cumulative Energy Distribution}

\begin{figure*}
    \centering
    \includegraphics[width=0.8\linewidth]{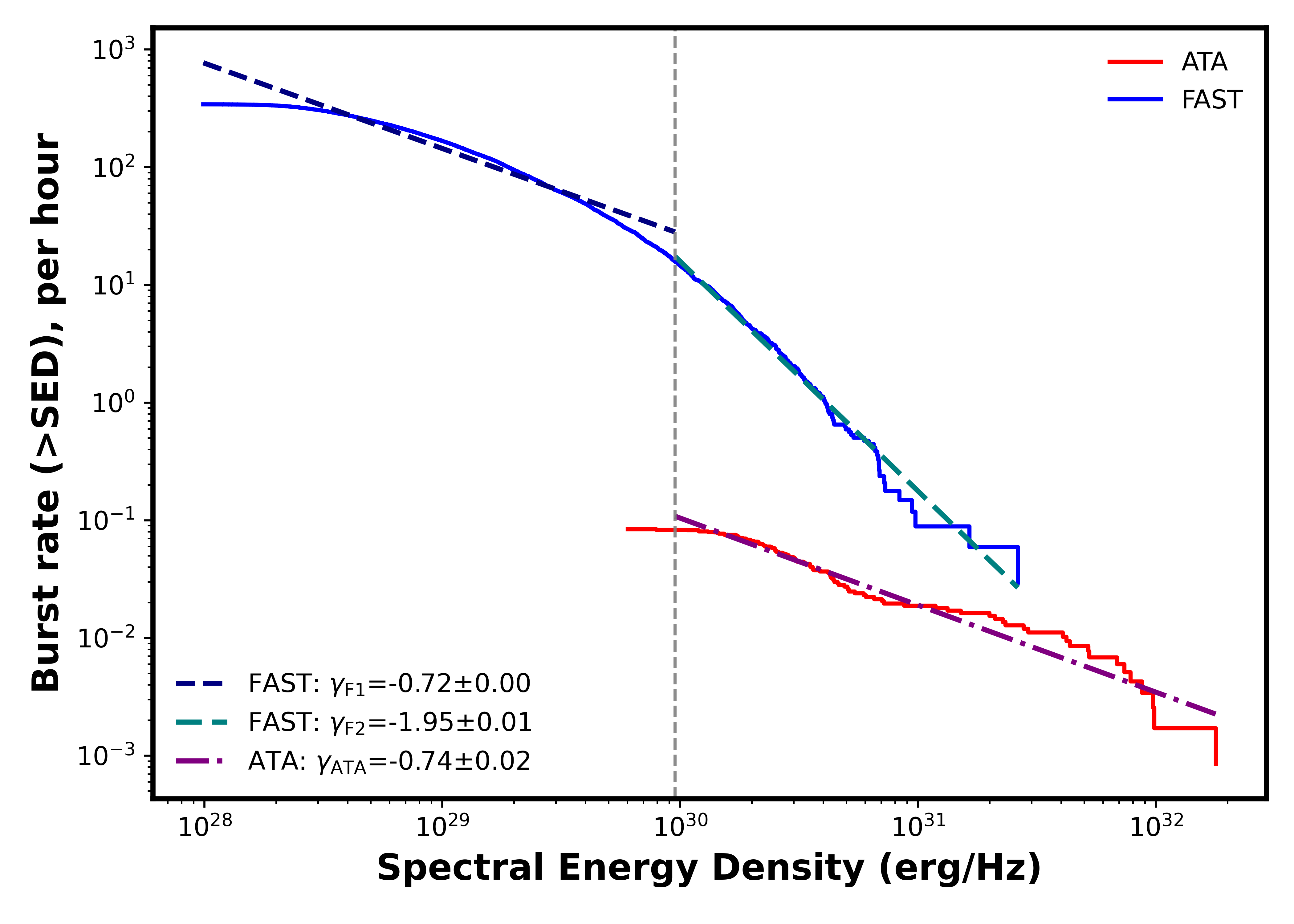}
    \caption{Log-scale plot of the cumulative burst rate as a function of spectral energy density (SED) for the ATA bursts, compared with the FAST energetics from \citet{FAST_main_paper}. The ATA distribution is fit with a power law of index $\gamma = -0.74 \pm 0.02$ above the completeness limit. The dashed vertical line marks the ATA completeness limit ($9.5 \times 10^{29}$\,erg\,Hz$^{-1}$), which is also taken as the break point of the FAST power-law fit. The significantly shallower ATA slope reflects the bursts missed while the ATA was not observing during the FAST high-activity period around MJD\,60380.}
    \label{fig:cummulative_energy_FAST_ATA}
\end{figure*}

Figure~\ref{fig:cummulative_energy_FAST_ATA} presents the cumulative burst rate (bursts per hour) as a function of spectral energy density (SED = isotropic burst energy / bandwidth) for both our ATA sample and the FAST sample of \citet{FAST_main_paper}. Above a well-defined ATA completeness threshold of $9.5 \times 10^{29}$\,erg\,Hz$^{-1}$ (vertical dashed line, corresponding to a minimum fluence of 2\,Jy\,ms), the ATA distribution follows a power law,
\begin{equation}
N(>SED) \propto SED^{\gamma},
\end{equation}
with a best-fit index of $\gamma = -0.74 \pm 0.02$, obtained from a least-squares fit in log space using \texttt{scipy.optimize.curve\_fit}. This relatively shallow slope indicates that high-energy bursts contribute significantly to the total energy output, in contrast to a steep distribution dominated by low-energy events. The deviation from power-law behavior below the completeness limit is the expected consequence of observational bias, as low-energy bursts fall below the detection threshold. Power-law energy distributions are commonly reported for repeating FRBs, and our shallow high-energy behavior echoes the flattened high-energy tail found for FRB\,20240114A by \citet{energydist_paper}, who note that such a tail resembles the distributions of apparent non-repeaters and may point to a common origin between the two classes. We caution that our SED index is not directly comparable to the differential energy index of $\gamma=-1.20$ measured at 2.25\,GHz by \citet{shangai_longterm_paper}, since these refer to different quantities and frequency ranges.

For the FAST comparison shown in the same figure, we fit a power law to the FAST data with a break fixed at the ATA completeness threshold. The slopes of the FAST and ATA fits differ significantly. We attribute this to the fact that the ATA was not observing during the very high burst-rate active period seen by FAST around MJD\,60380 (11 March 2024); we therefore likely missed many low-frequency, low-energy bursts, which accounts for the difference in activity at comparable SED.

The high-energy behavior of FRB\,20240114A can also be placed in the broader context of energetics studies of other active repeaters. A compilation of cumulative burst-energy distributions for several active repeating FRBs is presented in Table 2 of \cite{Ujjwal_GMRT_paper}, including FRB\,20121102A \citep{{Hewitt2022},Aggarwal_2021,Cruces2021,Wang2019}, FRB\,20200120E \citep{Nimmo2023}, FRB\,20201124A \citep{{Zhang2022},Xu2022}, FRB\,20220912A \citep{{Konijn2024},Zhang2023}, and the FRB\,20240114A. These studies generally find that burst energies follow power-law distributions, although the measured slopes span a wide range and often depend on observing frequency, activity state, and the adopted completeness threshold. The cumulative distributions typically exhibit indices between $\sim -0.4$ and $-2.3$, with evidence in several sources for a break or flattening at high energies. In this context, the ATA value of $\gamma=-0.74\pm0.02$ lies toward the shallower end of the observed range and is broadly consistent with a population in which energetic bursts contribute substantially to the overall burst-energy budget.

Recently, \cite{HyperFlash2026} reported the discovery of some of the most energetic radio bursts yet observed from a repeating FRB, from the FRB\,20240114A. These events reach isotropic-equivalent energies substantially exceeding the typical burst population and extend the known energy distribution of the source by more than an order of magnitude. The HyperFlash study argues that the highest-energy events constitute a rare but important component of the overall burst population and may require a separate statistical description from ordinary bursts. Although our ATA sample does not include bursts reaching the extreme HyperFlash energies, the observed shallow cumulative slope implies a relatively enhanced occurrence rate of high-energy events compared to a steep power-law distribution. Consequently, our results are qualitatively consistent with the emerging picture that FRB\,20240114A possesses an unusually extended high-energy tail that becomes increasingly apparent when large observing campaigns are combined across multiple facilities across multiple frequencies. 

\subsection{Normalized Drift Rate versus Duration}

\begin{figure*}
    \centering
    \includegraphics[width=0.8\linewidth]{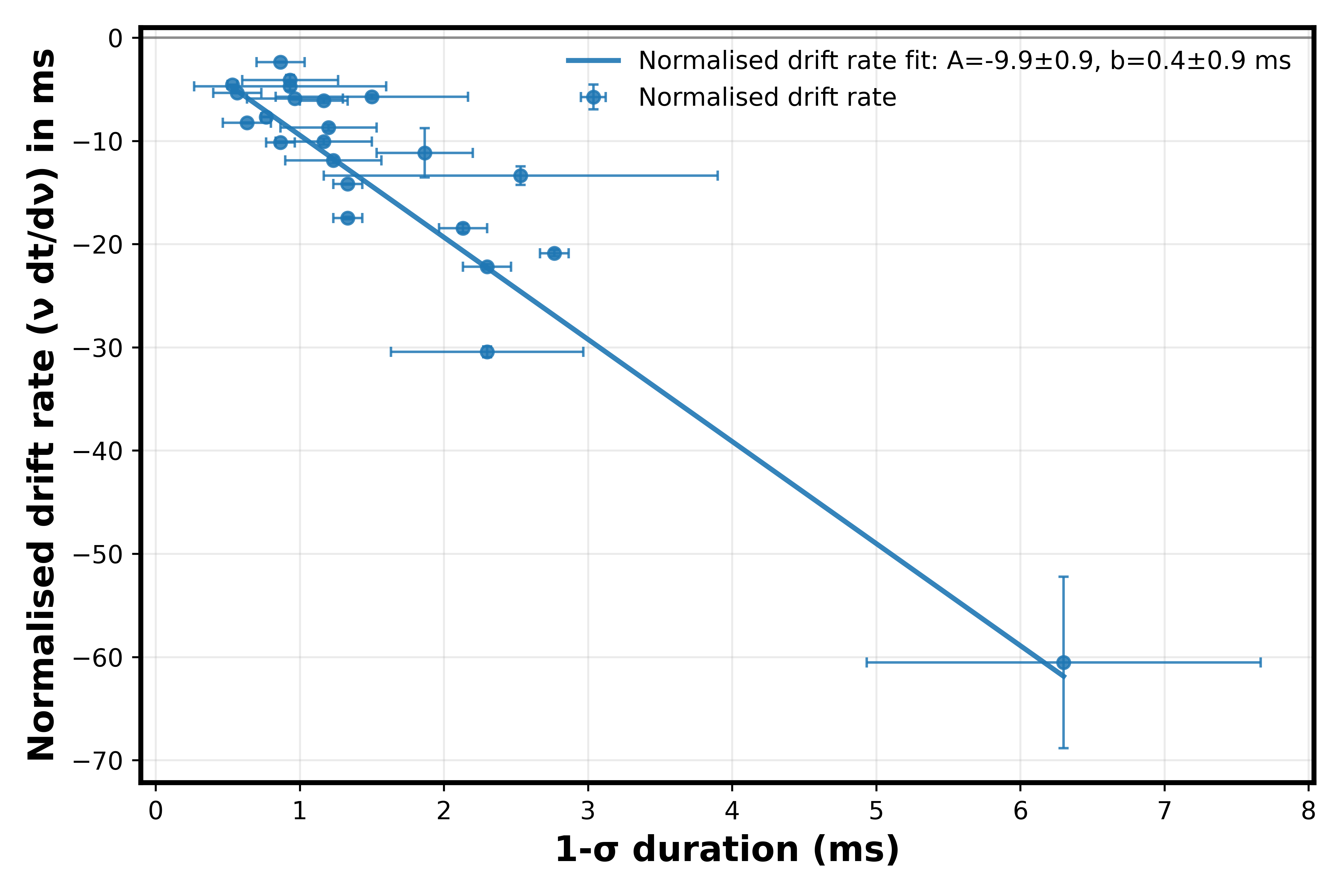}
    \caption{Normalised Drift Rate vs Duration using multi-component drift rates. The solid line is a fit of the form $\nu \text{d}t/\text{d}\nu=A\sigma_t+b$, fitted using the orthogonal distance regression method, via the \texttt{scipy.odr} package. The values obtained for $A$ and $b$ are consistent with the values found for other FRB repeaters as well as for the analogous relationship found between the sub-burst slope (or intra-burst drift) and duration.}
    \label{fig:drift_rate_vs_duration}
\end{figure*}

Another spectro-temporal relationship that can be characterized for FRB repeaters is the drift rate as a function of burst duration. Figure \ref{fig:drift_rate_vs_duration} shows the relationship between the normalized drift rate (center frequency/drift rate) and the burst duration (expressed as the 1-sigma temporal width). The relationship between these two properties appears linear, a relationship also observed by \cite{Chamma2025} in the drift rates from the repeaters FRB 20121102A, FRB 20220912A, FRB 20200120E, and others. To quantify this relationship and to compare with earlier results, we fit a line of the form $\nu\cdot \text{d}t/\text{d}\nu = A\sigma_t+b$ to our multi-component drift rate measurements. We obtain $A=-9.9(9)$ and $b=0.4(9)$ ms, which describe the data well. The triggered relativistic dynamical model (TRDM; \citealt{Rajabi2020}) predicts an analogous relationship but between the sub-burst slope (or intra-burst drift) and duration of individually resolved FRB pulses. The sub-burst slope relation is observed in multiple repeaters, for e.g. \citealt{Chamma2021,Brown2024,Jahns2023}. \cite{Chamma2025} observe that the sub-burst slope relation also well describes the analogous multi-component drift rate relation with duration. Their reported fit values found for sub-burst slopes measured from the repeater FRB 20121102A are $A=-8.6(4)$ and $b=2(1)\times10^{-2}$ ms. \cite{Jahns2023} did a similar analysis and found an $A$ value of approximately $-12.5$ after converting to the present formalism \citep{Chamma2025}. These values are all similar to one another, and our results for FRB 20240114A are consistent with earlier suggestions that the scaling of this relation is the same not only between different repeating FRBs but also between sub-burst slopes and multi-component drift rates. In the TRDM, the absolute value of A in the sub-burst slope relation is the ratio between the proper delay time (between trigger and emission) and the proper duration of emission \citep{Rajabi2020}. However, it is unclear how this parameter should be interpreted in the context of multi-component drift rates.

The TRDM makes other predictions about the relationships already discussed and may most notably be applicable to the drift rate vs frequency relationship discussed in Section \ref{subsec:driftvfreq}. Specifically, the sharp increase in the drift rate magnitude observed at 4000 MHz is expected by eq. 10 of \cite{Chamma_2023_FRBGUI}, where the drift rate magnitude is expected to increase quadratically with the center frequency (i.e. $\text{d}\nu/\text{d}t \propto \nu^2$). Generally speaking, the drift-duration and drift-frequency relationships discussed here are consistent with a narrowband emission process and dynamical motions in the source that transform the observed FRB signal.

\section{Conclusion}
\label{sec:conclusion}

We have presented a high-frequency, wideband study of the hyperactive repeating source FRB\,20240114A \citep{CHIME_FRB_2025} with the ATA. Over a 1167\,hr campaign spanning 27 January to 29 October 2024, we stepped ten LO tunings from $\sim$900\,MHz to 7620\,MHz and detected 97 bursts between $\sim$900\,MHz and $\sim$5\,GHz at a structure-optimized DM of 528.1\,pc\,cm$^{-3}$, extending our first report of wideband emission above 2\,GHz \citep{ATA_Atel}. Using \texttt{FRBGUI} \citep{Chamma_2023_FRBGUI}, we measured the spectro-temporal and energy properties of these bursts and their 162 sub-components, and compared them with the contemporaneous Parkes/Murriyang \citep{Uttarkar2026}, Effelsberg \citep{limaye2026}, FAST \citep{FAST_main_paper}, and Tianma \citep{shangai_longterm_paper} campaigns. Our main findings are:

\begin{itemize}
    \item The burst activity is strongly frequency- and time-dependent, peaking at a rate of $0.25$\,hr$^{-1}$ in the 2244--3588\,MHz band around MJD\,60500; no bursts were detected in $\sim$305\,hr spanning 4932--7620\,MHz, confirming that the emission, while broadband, remains band-limited in any single epoch \citep{limaye2026,shangai_longterm_paper}, as seen in other repeaters \citep{pleunis_lofar_2021}.
    \item The ATA captured a dense, high-frequency (2.4--4.6\,GHz) burst storm near MJD\,60480--60546 that was largely missed by other campaigns. This storm occurs at a phase where the claimed $\sim$112.9\,day central-frequency periodicity of \citet{Li2026_freqperiod} predicts low-frequency emission, in tension with a strictly phase-coherent modulation and underscoring the role of selection effects in periodicity claims such as the $\sim$143.4\,day activity cycle reported by \citet{FAST_periodicity_paper2}.
    \item The absolute burst bandwidth and drift-rate magnitude increase with center frequency, while the burst duration decreases, in agreement with the Effelsberg \citep{limaye2026}, Tianma \citep{shangai_longterm_paper}, and FRB\,20121102A \citep{gajjar_highest_2018} results; the bursts drift predominantly downward (the ``sad-trombone'' effect), consistent with the FAST burst-cluster analysis of \citet{FAST_drifting_paper3}.
    \item The fractional bandwidth is approximately scale-invariant with frequency; our 3-$\sigma$ median ($\sim$35\%) exceeds the $\sim$10\% reported by \citet{limaye2026} and \citet{energydist_paper}, an offset we attribute to our use of 3-$\sigma$ widths and to the ATA's sensitivity being limited to the brightest, preferentially broadband bursts, in contrast to the deeper samples of repeaters that typically occupy $\sim$10--30\% \citep{pleunis_lofar_2021}.
    \item The cumulative spectral-energy-density distribution follows a shallow power law ($\gamma = -0.74 \pm 0.02$) above the completeness limit, consistent with the flattened high-energy tail reported by \citet{energydist_paper}, while the normalized drift rate scales linearly with burst duration ($A=-9.9(9)$), in line with the shared spectro-temporal relation seen across repeating FRBs \citep{Chamma2021,Chamma2025,Rajabi2020}.
\end{itemize}

Taken together, our results establish FRB\,20240114A as a genuinely broadband, hyperactive repeater whose spectral activity migrates to higher frequencies at later epochs \citep{Uttarkar2026}, and demonstrate the unique value of sustained, wideband monitoring above 2\,GHz. The ATA detection of a high-frequency burst storm unseen by other facilities highlights that the apparent periodicities and chromatic behavior reported for this source \citep{Li2026_freqperiod,FAST_periodicity_paper2} can be strongly shaped by each instrument's bandwidth and sensitivity. Disentangling intrinsic emission physics from propagation and selection effects will require continued multi-telescope, high-cadence campaigns with overlapping wideband coverage over many more activity cycles.

\begin{acknowledgements}
The Allen Telescope Array (ATA) is owned and operated by the SETI Institute. The ATA refurbishment program and its ongoing operations have received substantial support from Franklin Antonio. Breakthrough Listen is managed by the Breakthrough Initiatives, sponsored by the Breakthrough Prize Foundation. 
\end{acknowledgements}

\section*{Data Availability}
The full table of spectro-temporal properties for all 97 bursts is available as a CSV file at \href{https://zenodo.org/uploads/19429571}{zenodo.org/uploads/19429571}. The dynamic spectra in archive 
or filterbank formats will be shared on reasonable request to the 
corresponding author.

\bibliography{references_in_use}
\bibliographystyle{aasjournalv7}

\end{document}